\documentclass{emulateapj}

\usepackage{amsmath}

\usepackage{graphics}

\newcommand{\bd}{\begin{displaymath}}
\newcommand{\ed}{\end{displaymath}}

\slugcomment{Received 2009 December 1; accepted 2010 May 3}

\shorttitle{Growth of massive black holes at their late stage}
\shortauthors{Y-D Xu \& X Cao }

\begin{document}

\title{Growth of massive black holes at their late stage}

\author{Ya-Di Xu$^{1}$, Xinwu Cao$^{2}$}
\affil{$^{1}$Physics Department, Shanghai Jiao Tong University, 800
Dongchuan Road, Shanghai 200240, China}

\affil{$^{2}$Key Laboratory for Research in Galaxies and Cosmology,
Shanghai Astronomical Observatory, Chinese Academy of Sciences, 80
Nandan Road, Shanghai 200030, China\\
Email: ydxu@sjtu.edu.cn, cxw@shao.ac.cn}

\clearpage

\begin{abstract}

It is believed that the local massive black holes were dominantly
grown up through accretion during quasar phases, while a fraction of
the local black hole mass was accumulated through accreting gases at
very low rates. We derive the black hole mass density as a function
of redshift with the bolometric luminosity function of active
galactic nuclei (AGN) assuming that massive black holes grew via
accreting the circumnuclear gases, in which the derived black hole
mass density is required to match the measured local black hole mass
density at $z=0$. Advection dominated accretion flows (ADAFs) are
supposed to present in low luminosity active galactic nuclei
(AGNs)/normal galaxies, which are very hot and radiate mostly in the
hard X-ray band. Most of the X-ray background (XRB) is contributed
by bright AGNs, and a variety of AGN population synthesis models
were developed to model the observed XRB in the last two decades.
Based on our derived black hole mass density, we calculate the
contribution to the XRB from the ADAFs in faint AGNs/normal galaxies
with a given Eddington ratio distribution, which is mostly in hard
X-ray energy band with an energy peak at $\sim 200$~keV. The growth
of massive black holes during ADAF phase can therefore be
constrained with the observed XRB. Combining an AGN population
synthesis model with our results, {we find that the fitting on the
observed XRB, especially at hard X-ray energy band with $\ga
100$~keV, is improved provided the contribution of the ADAFs in
low-luminosity AGNs/normal galaxies is properly included.} It is
found that less than $\sim$15 per cent of local massive black hole
mass density was accreted during ADAF phases. {We suggest that more
accurate measurements of the XRB in the energy band with $\ga
100$~keV in the future may help constrain the growth of massive
black holes at their late stage.} We also calculate their
contribution to the extragalactic $\gamma$-ray background (EGRB),
and find that less than $\sim1\%$ of the observed EGRB is
contributed by the ADAFs in these faint sources.
\end{abstract}


\section{Introduction}

It is well believed that almost all galaxies contain massive black
holes at their centers, and a tight correlation was revealed between
central massive black hole mass and the velocity dispersion of the
galaxy \citep[][]{f00,g00}. The growth of massive black holes at the
centers of galaxies may probably be linked to accretion processes
\citep{s82}. \citet{y02} estimated the black hole masses from the
stellar velocity dispersions of galaxies measured by the Sloan
Digital Sky Survey (SDSS) using the empirical relation between black
hole mass and the velocity dispersion, and the local black hole mass
density was derived. They further calculated the black hole mass
density accreted during optical bright quasar phases using an
optical quasar luminosity function (LF), and found that the accreted
mass density is consistent with the local black hole mass density
estimated from the velocity dispersions, if a radiative efficiency
$\sim 0.1$ is adopted for quasars \citep*[also
see][]{ma04,shankar04,h07,shankar09}. This implies that the growth
of massive black holes through accretion during optically bright
quasar phases may probably be important, if a radiative efficiency
$\sim 0.1$ is adopted. If the massive black holes are spinning
rapidly, their radiative efficiency can be higher than that they
adopted, the black hole mass density accreted during quasar phases
would be lower than the measured local black hole mass density
\citep*[e.g.,][]{c07,cl08,l10}. Thus, one cannot neglect the
contribution from accretion at low rates to the growth of massive
black holes, provided the duration of accretion at low rates is as
long as Hubble timescale.

The cosmological X-ray background (XRB) is mostly contributed by
AGNs \citep{h98,s98}, which can be used to constrain massive black
hole accretion history \citep*[e.g.,][]{e02}. In the most popular
synthesis models of the XRB based on the unification schemes for
AGNs, the cosmological XRB contributed by Compton-thin AGNs can
account for $\sim 80$\% of the observed XRB
\citep*[e.g.,][]{u03,g07}. The residual XRB can be explained
provided the same number of Compton-thick AGNs with $\log N_{\rm
H}=24-25$ as those with $\log N_{\rm H}=23-24$ is included
\citep{u03}. \citet{d97} have alternatively proposed that the hard
XRB above 10 keV may be dominated by the thermal bremsstrahlung
emission from the advection dominated accretion flows (ADAFs) in
low-luminosity AGNs. {Due to the difficulties on detecting
Compton-thick AGNs, the space number density of Compton-thick AGNs
is still quite debated \citep*[e.g.,][]{u03,tu05,g07,tuv09}. Based
on the surveys with \textit{Swift} and \textit{INTEGRAL} satellites,
\citet{tuv09} suggested that the space number density of
Compton-thick AGNs may be significantly lower than those adopted in
the previous AGN population synthesis models to explain the observed
XRB.} Due to the uncertainty of the number density of Compton-thick
AGNs, the residual XRB could be attributed to both the Compton-thick
AGNs and the ADAFs in faint AGNs and normal galaxies. Therefore, the
observed XRB can be used to constrain massive black hole accretion
history at their late stage when they are accreting at low rates
\citep{c05,c07}. \citet{c07} calculated the contribution to the XRB
from the ADAFs in faint AGNs/normal galaxies, and compared it with
the residual XRB. Their results showed that less than $\sim$5 per
cent of local massive black hole mass density was accreted during
ADAF phases, otherwise the XRB contributed from ADAFs will surpass
the observed residual XRB even if no Compton-thick AGNs are
included.  For simplicity, they adopted an average mass accretion
rate for faint AGNs/normal galaxies in calculating the contribution
of ADAFs to the XRB. The radiative efficiency of standard thin
accretion disks in bright AGNs does not vary with mass accretion
rate. However, the faint AGNs/normal galaxies may probably contain
ADAFs, of which the radiative efficiencies vary with mass accretion
rate \citep{n95}. The average radiative efficiency for a population
of sources containing ADAFs accreting at different rates can be
calculated by weighing over the distribution of mass accretion rate.
The Eddington ratios of AGNs spread over several orders of magnitude
\citep*[e.g.,][]{h02,h06,cx07}.
The Eddington ratio distribution for accreting massive black hole
holes derived from observations exhibits nearly a power-law
distribution with an exponential cutoff at high Eddington ratio
\citep{m08,h09,kh09}. In this paper, we will adopt the Eddington
ratio distribution of AGNs derived by \citet{h09} to calculate the
contribution of faint AGNs/normal galaxies to the XRB. Compared with
the residual XRB,
the constraints on the fraction of local black
hole mass accreted in ADAF phases are derived. The cosmological
parameters $\Omega_{\rm M}=0.3$, $\Omega_{\Lambda}=0.7$, and
$H_0=70~ {\rm km~s^{-1}~Mpc^{-1}}$ have been adopted in this work.


\section{Black hole mass functions}

In this section, we derive the black hole mass densities as
functions of redshift in a similar way as that done in \citet{c07}.
In this work, we use the bolometric quasar luminosity function (QLF)
derived by \citet{h07} to calculate the black hole mass densities.
We summarize our calculations as follows \citep*[see][for the
details]{c07}.

\citet{h07}'s QLF is calculated by using a large set of observed
quasar luminosity functions in various wavebands, from the IR
through optical, soft and hard X-rays \citep*[see][for the
details]{h07},
\begin{equation}
\frac{{\rm d} \Phi(L,z)}{{\rm d}\log L}=\frac
{\phi_{*}}{(L/L_{*})^{\gamma_{1}}+(L/L_{*})^{\gamma_{2}}},
\label{hxlf}
\end{equation}
with normalization $\phi_{*}$,  break luminosity $L_{*}$, faint-end
slope $\gamma_{1}$, and bright-end slope $\gamma_{2}$. The break
luminosity $L_{*}$ evolves with the redshift is given by
\begin{equation}
\log L_{*}=(\log
L_{*})_{0}+k_{L,1}\xi+k_{L,2}\xi^{2}+k_{L,3}\xi^{3},
\end{equation}
and the two slopes $\gamma_{1}$  and $\gamma_{2}$ evolves with
redshift as
\begin{equation}
\gamma_{1}=(\gamma_{1})_{0}(\frac{1+z}{1+z_{\rm
ref}})^{k_{\gamma_{1}}},
\end{equation}
and
\begin{equation}
\gamma_{2}=\frac{2(\gamma_{2})_{0}}{(\frac{1+z}{1+z_{\rm
ref}})^{k_{\gamma_{2},1}}+(\frac{1+z}{1+z_{\rm
ref}})^{k_{\gamma_{2},2}}}.
\end{equation}
The parameter $\xi$ is
\begin{equation}
\xi=\log(\frac{1+z}{1+z_{\rm ref}}),
\end{equation}
and $z_{\rm ref}=2$ is fixed.  The best-fit parameters we adopted in
this work are as follows: $\log \phi_{*}({\rm Mpc}^{-3})=-4.825\pm
0.060$, $[\log L_{*}(3.9\times 10^{33} {\rm
erg~s}^{-1})]_{0}=13.036\pm 0.043$, $k_{L,1}=0.632\pm 0.077$,
$k_{L,2}=-11.76\pm 0.38$, $k_{L,3}=-14.25\pm 0.80$,
$(\gamma_{1})_{0}=0.417\pm 0.055$, $k_{\gamma_{1}}=-0.623\pm 0.132$,
$(\gamma_{2})_{0}=2.174\pm 0.055$, ${k_{\gamma_{2},1}}=1.460\pm
0.096$, and ${k_{\gamma_{2},2}}=-0.793\pm 0.057$ \citep{h07}.

In this work, all the sources described by this QLF $\Phi(L,z)$ are
referred as active galaxies, {which are supposed to have mass
accretion rates $\dot{m}\gtrsim \dot{m}_{\rm crit}=0.01$ [defined as
$\dot{m}=0.1\dot{M}c^2/L_{\rm Edd}$, and $L_{\rm Edd}=1.3\times
10^{38}~{\rm erg~s}^{-1} (M_{\rm bh}/{\rm M_\odot})$] and contain
standard radiative efficient accretion disks.} The cosmological
evolution of black hole mass density caused by accretion during
active galaxy phases is described by
\begin{equation}
{\frac {{\rm d}\rho_{\rm bh}^{\rm A}(z)}{{\rm d}z} }= {\frac {{\rm
d}t}{{\rm d}z}}{1\over {\rm M_\odot}} \int\limits {\frac
{(1-\epsilon)L}{\epsilon c^2}} \frac{{\rm d} \Phi (L, z)}{{\rm d}
\log L} {\rm d} \log L,~~~ \label{rhobha}
\end{equation}
where $\Phi (L, z)$ is the bolometric QLF given by \citet{h07},
$\rho_{\rm bh}^{\rm A}(z)$ (in units of ${\rm M_\odot~Mpc^{-3}}$) is
the black hole mass density accreted during active galaxy phases
from $z_{\rm max}$ to $z$, and $\epsilon$ is the average radiative
efficiency for active galaxies.

For those inactive galaxies with $L<10^{41}$ ergs ${{\rm s}^{-1}}$,
their mass accretion rates are very low. {The term ``inactive
galaxies" used in this work does not mean that they are really
inactive. Compared with active galaxies, the massive black holes in
inactive galaxies are still accreting, but at low rates with
$\dot{m}\lesssim \dot{m}_{\rm crit}=0.01$,} and {radiatively
inefficient} ADAFs are suggested to be present in inactive galaxies
\citep*[e.g.,][]{n02}. The black hole mass density $\rho_{\rm
bh}^{\rm B}(z)$ accreted during inactive galaxy phases between $z$
and $z_{\rm max}$ can be calculated by
\begin{equation}
{\frac {{\rm d} \rho_{\rm bh}^{\rm B}(z)}{{\rm d}z}} ={\frac
{\rho_{\rm bh}^{\rm inact}(z)\dot{m}_{\rm inact}^{\rm
aver}L_{\odot,\rm Edd} (1-\epsilon^{\rm ADAF})} {0.1{\rm M}_\odot
c^2}} {\frac {{\rm d}t}{{\rm d}z}}, ~~~ \label{rhobhb}
\end{equation}
where  $\dot{m}_{\rm inact}^{\rm aver}$ is the average dimensionless
mass accretion rate for the inactive galaxies, $\rho_{\rm bh}^{\rm
inact}(z)$ is the black hole mass density of inactive galaxies at
redshift $z$, $\epsilon^{\rm ADAF}$ is the average radiative
efficiency of the ADAFs in those inactive galaxies.  The radiative
efficiency $\epsilon^{\rm ADAF}$ is usually much lower than 0.01, so
we approximately adopt $\epsilon^{\rm ADAF}\simeq 0$ in our
calculations on black hole mass densities.

Assuming that the growth of massive black hole is dominated by
accretion, the total black hole mass density is given by
\begin{equation}
\rho_{\rm bh}(z)\simeq \rho_{\rm bh}^{\rm acc}(z)+\rho_{\rm
bh}(z_{\rm max}) =\rho_{\rm bh}^{\rm A}(z)+\rho_{\rm bh}^{\rm
B}(z)+\rho_{\rm bh}(z_{\rm max}), \label{rhobhtotb}
\end{equation}
where $\rho_{\rm bh}(z_{\rm max})$ is the total black hole mass
density at $z_{\rm max}$. For active galaxies, the black hole mass
density at redshift $z$ can be calculated from the QLF by
\begin{equation} \rho^{\rm act}_{\rm
bh}(z)={\frac {1}{\lambda^{\rm aver}_{\rm act}L_{\rm Edd,\odot}}}
\int\limits L\frac{{\rm d} \Phi (L, z)}{{\rm d} \log L} {\rm d} \log
L~~~ {\rm M}_{\odot} {\rm Mpc^{-3}}, \label{bhmdensact}
\end{equation}
where $\Phi (L, z)$ is the bolometric QLF given by Eq. (\ref{hxlf}),
{and $\lambda^{\rm aver}_{\rm act}=L_{\rm bol}/L_{\rm Edd}$ is the
average Eddington ratio for active galaxies.} In this work, we adopt
$\lambda^{\rm aver}_{\rm act}=0.25$ as that derived from a large
bright AGN sample \citep{k06}. The black hole mass density for
inactive galaxies $\rho_{\rm bh}^{\rm inact}(z)$ can be calculated
with
\begin{equation}
\rho_{\rm bh}^{\rm inact}(z)=\rho_{\rm bh}(z)-\rho_{\rm bh}^{\rm
act}(z) =\rho_{\rm bh}^{\rm A}(z)+\rho_{\rm bh}^{\rm B}(z)+\rho_{\rm
bh}(z_{\rm max})-\rho_{\rm bh}^{\rm act}(z). \label{rhobhinact}
\end{equation}

{We assume $\rho_{\rm bh}^{\rm act}(z)=0.5\rho_{\rm bh}(z_{\rm
max})$ at $z=z_{\rm max}$, i.e., a half of all massive black holes
are active at $z_{\rm max}$, as that used in \citet{ma04}, and the
black hole mass density $\rho_{\rm bh}(z_{\rm max})=2\rho_{\rm
bh}^{\rm act}(z_{\rm max})$ is obtained from Eq.
(\ref{bhmdensact}).} Thus, the black hole mass density as function
of redshift $z$ can finally be calculated by integrating Eqs.
(\ref{rhobha}) and (\ref{rhobhb}) from $z_{\max}$ to $z$ with Eqs.
(\ref{rhobhtotb})-(\ref{rhobhinact}), when the three parameters,
$\lambda_{\rm act}^{\rm aver}$, $\dot{m}_{\rm inact}^{\rm aver}$ and
$\epsilon$ are specified. We adopt $z_{\rm max}=5$ in all
calculations, and the value of $\epsilon$ is tuned to make the
derived total black hole mass density $\rho_{\rm bh}(z)$ match the
local black hole mass density $\rho_{\rm bh}^{\rm
local}(z)=4.0\times 10^{5}{\rm M_\odot~Mpc^{-3}}$ at $z=0$
\citep*[e.g.,][]{s99,ma04}, when the value of $\dot{m}_{\rm
inact}^{\rm aver}$ is specified. {The black hole mass accreted
during $z\ge 5$, $\rho_{\rm bh}(z=5)$, should be $\ll\rho_{\rm
bh}(0)$, which implies that it will make little difference on our
calculations even if a better QLF is available for $z_{\rm max}>5$.
Thus, we will not extrapolate the present QLF to higher redshifts.}

\section{Eddington Ratio distribution for inactive galaxies}




The observed Eddington ratio distribution can be described by
\begin{equation}
f_{\lambda}(\lambda)=\frac{{\rm d}N}{N{\rm d}\log
\lambda}=C_{0}\left
(\frac{\lambda}{\eta}\right)^{-\kappa}\exp\left(-{\lambda\over\eta}\right
), \label{edd_ratio_dis}
\end{equation}
where $\lambda$ is the Eddington ratio, $\lambda\equiv L/{L_{\rm
Edd}}$, and $C_{0}$ is the normalization \citep{h09}. They suggested
that $\kappa\approx 0.3-0.8$, and $\eta=0.2-0.4$.
This Eddington ratio distribution is consistent with the
self-regulated black hole growth model, in which feedback produces a
self-regulating "decay" or "blowout" phase after the AGN reaches
some peak luminosity and begins to expel gas and shut down accretion
\citep*{h05a,h05b,h09}.

In this work, we assume the distribution of dimensionless mass
accretion rate have a similar form as Eq. (\ref{edd_ratio_dis}),
\begin{equation}
f_{\rm m}(\dot{m})=\frac{{\rm d}N}{N{\rm d}\log \dot{m}}=C_{1}\left
(\frac{\dot{m}}{\eta}\right )^{-\kappa_{\rm m}}\exp\left
(-{\dot{m}\over\eta}\right )\simeq C_{2}{\dot{m}}^{-\kappa_{\rm m}},
\label{mdot_dis}
\end{equation}
for low-luminosity AGNs with $\dot{m}<\dot{m}_{\rm crit}$, where
$C_{2}$ is the normalization. {In all our calculations, we drop the
exponential term in (\ref{mdot_dis}), which is a good approximation
because $\dot{m}<\dot{m}_{\rm crit}<<\eta$, and the accretion rate
distribution (\ref{mdot_dis}) is always normalized by assuming all
inactive black holes to be accreting with rates in the range of
$\dot{m}_{\rm min}\le\dot{m}\le\dot{m}_{\rm max}=\dot{m}_{\rm
crit}$. Thus, the accretion rate distribution can be described by
one parameter $\kappa_{\rm m}$ with specified mass accretion rate
range for inactive galaxies.} The standard thin accretion disks are
present in bright quasars, while they will transit to ADAFs provided
the mass accretion rates are lower than the critical value
$\dot{m}_{\rm crit}$ \citep*[e.g.,][]{nmq98,n02}. The critical
dimensionless mass accretion rate $\dot{m}_{\rm crit}\simeq 0.01$ is
suggested either by observations or theoretical model calculations
\citep*[see][for a review and references therein]{n02}. For inactive
galaxies with $\dot{m}\la \dot{m}_{\rm crit}$, their average mass
accretion rate can be calculated with
\begin{equation}
\dot{m}_{\rm inact}^{\rm aver}={\int^{\dot{m}_{\rm
crit}}_{\dot{m}_{\rm min}}\dot{m}f_{\rm m}(\dot{m}){\rm d}\log
\dot{m}},
 \label{mdot_aver}
\end{equation}
where the minimum accretion rate, $\dot{m}_{\rm
min}=1.0\times10^{-5}$ is adopted in all our calculations. Thus, the
value of parameter $\kappa_{\rm m}$ corresponds to an average mass
accretion rate $\dot{m}_{\rm inact}^{\rm aver}$.

\section{Contributions of inactive galaxies to the cosmological background radiation}


 We employ the approach suggested by \citet{manmoto00} to calculate the
global structure of an ADAF surrounding a massive black hole in the
general relativistic frame. All the radiation processes are included
in the global structure calculations \citep*[see][for details and
the references therein]{manmoto00}. The global structure of an ADAF
surrounding a $10^8{\rm M_{\odot}}$ black hole with spin parameter
$a$ can be calculated, if the model parameters, { dimensionless mass
accretion rate $\dot{m}$, magnetic field strength relative to gas
pressure $\beta$, defined as $p_{\rm m}=B^2/8\pi=(1-\beta)p_{\rm
tot}$ ($p_{\rm tot}=p_{\rm gas}+p_{\rm m}$), the fraction of the
released gravitational energy directly heating the electrons
$\delta$, and the conventional viscosity parameter $\alpha$,} are
specified. The constraints on the values of these parameters were
discussed in \citet{c07}. We adopt the same values as that work,
i.e., $\alpha=0.2$, and $\beta=0.8$, with which no global solution
is available for $\dot{m}\ga 0.01$. This is consistent with the
observations \citep*[e.g.,][]{n02}. The value of $\delta$ is still a
controversial issue. In most of our calculations, we adopt a
conventional value of $\delta=0.1$ as that adopted \citet{c07}. We
also calculate the case with $\delta=0.01$ for comparison.


The inner region of the ADAF is very hot, and the temperature of
protons can be as high as $\sim 10^{12}\rm K$. Thus, $\gamma$-ray
emission may be produced through the pion production processes in
the proton-proton (p-p) collisions and subsequently decay of neutral
pions, which can be described by \citep{m97}
\begin{equation}
p+p\rightarrow~ p+p+\pi^{0}, ~\pi^{0}\rightarrow
~\gamma_{1}+\gamma_{2}.
\end{equation}
The production of gamma rays through p-p collisions has been studied
in several works \citep{d86,g82a,g82b}. \citet{s71} showed that the
gamma-ray spectrum produced in unit volume of the flow from
$\pi^{0}$ decay can be calculated by
\begin{equation}
L_{\nu}=h^2\nu~f_{\gamma}(E_{\gamma})=2h^2\nu~\int^{\infty}_{E_{\pi_{\rm
min}}}{\rm
d}E_{\pi}\frac{f_{\pi}(E_{\pi})}{\sqrt{E_{\pi}^2-m_{\pi}^2}}~~~~{\rm
GeV~s^{-1}~Hz^{-1}~cm^{-3}},
\end{equation}
where $E_{\pi}$ is the pion energy in GeV, $m_{\pi}$ is the mass of
the pion in ${\rm GeV~ c^{-2}}$, $f_{\pi}(E_{\pi})$ is the $\pi^{0}$
spectrum, and $E_{\pi_{\rm min}}$ is the minimum pion energy
required to produce a gamma-ray with energy $E_{\gamma}$, which is
described by
\begin{equation}
E_{\pi_{\rm min}}=E_{\gamma}+\frac{m_{\pi}^2}{4E_{\gamma}}.
\end{equation}
The gamma-ray spectrum of an ADAF mainly depends on the density and
temperature distributions of the ions. The $\pi^{0}$ spectrum is
given by \citep{d86},
 \bd f_{\pi}(E_{\pi})=\frac{cn_{\rm
p}^2}{4m_{\pi}\theta_{\rm p}K^2_{2}(1/\theta_{\rm
p})}\int^{\infty}_{1}\rm d \gamma_{\rm r}\frac{(\gamma_{\rm
r}^2-1)}{[2(\gamma_{\rm r}+1)]^{1/2}} \ed \bd
~~~~\times\int^{\zeta/m_{\pi}}_{1}\rm d
\gamma^{*}(\beta^{*}\gamma^{*})^{-1}\frac{\rm
d\sigma^{*}(\gamma^{*};\gamma_{\rm r})}{\rm d \gamma^{*}} \ed \bd
~~~~~~~~~~~~~~~~~~~~~~~~~~\times
\{\exp[-q\gamma\gamma^{*}(1-\beta\beta^{*})]-\exp[-q\gamma\gamma^{*}(1+\beta\beta^{*})]\}
\ed
\begin{equation}
~~~~~~~~~~~~~~~~~~~~~~~~~~~~~~~~~{\rm cm^{-3}~s^{-1}~GeV^{-1}},
\end{equation}
where $n_{\rm p}$ is the number density of ions, $\theta_{\rm
p}=kT_{\rm i}/m_{\rm p}c^2$ is the dimensionless ion temperature,
$E_{\pi}=\gamma m_{\pi}$ is the pion energy in the observer's frame,
and { $K_{2}(x)$} is the modified Bessel function of order 2. The
differential cross section for the production of a neutral pion with
Lorentz factor $\gamma^{*}$ in the center-of-momentum system (CM) of
two colliding protons with relative Lorentz factor $\gamma_{\rm r}$
is denoted by ${\rm d\sigma^{*}(\gamma^{*};\gamma_{\rm r})}/{\rm d
\gamma^{*}}$, which is given in \citet{s71}. The quantities $q$ and
$\zeta$ { are} defined as $q=[2(\gamma+1)]^{1/2}/\theta_{\rm p}$ and
$\zeta=(S-4m_{\rm p}^2+m_{\pi}^2)/2S^{1/2}$, where $S=2m_{\rm
p}^2(\gamma_{\rm r}+1)$ \citep*[see][for the details]{s71}. Thus,
both the X-ray and $\gamma$-ray spectra of an ADAF can be calculated
when the global structure of the ADAF is available.


{We perform a set of spectral calculations for ADAFs surrounding
black holes with different masses, and find that the dependence of
ADAF spectrum on black hole mass is almost perfectly linear in
X-ray/$\gamma$-ray bands for $M_{\rm bh}\ga 10^6{\rm M}_\odot$,
provided all other parameters are fixed. Therefore we simply use the
spectrum $l_{E}(\dot{m})$ of an ADAF around a typical massive black
hole, $10^8{\rm M}_\odot$, accreting at the rate $\dot{m}$ as a
template spectrum to calculate the contribution of inactive galaxies
in unit of co-moving volume to the XRB by multiplying $\rho_{\rm
bh}^{\rm inact}(z)/10^{8}\rm M_{\odot}$ with the accretion rate
distribution (\ref{mdot_dis}) (see Eq. \ref{xrb} in this section).}
We can calculate the average spectrum of a population of ADAFs
accreting at rates with a distribution given by (\ref{mdot_dis}),
 \begin{equation} L_{\rm E}( E)=\int^{\dot{m}_{\rm crit}}_{\dot{m}_{\rm min}}f_{\rm m}(\dot{m})l_{E}(\dot{m}){\rm
d}\log\dot{m},
 \end{equation}
where $l_{E}(\dot{m})$ is X-ray and $\gamma$-ray spectrum from an
ADAF surrounding a $10^{8}\rm M_{\odot}$ black hole accreting with
$\dot{m}$.




The contribution of the ADAFs in all inactive galaxies to the
cosmological background radiation can be calculated by
\begin{equation}
f(E)=\frac{1}{10^{8}\rm M_{\odot}}\int^{z_{\rm
max}}_{0}\frac{\rho^{\rm inact}_{\rm bh}(z)(1+z)L_{\rm
E}[(1+z)E]}{4\pi~d^{2}_{\rm L}}\frac{{\rm d}V}{{\rm d}z}{\rm d}z,
\label{xrb}
\end{equation}
where $L_{\rm E}(E)$ is the template spectrum of ADAFs surrounding
$10^{8}\rm M_{\odot}$ black holes averaged over accretion rate
$\dot{m}$.

\section{Results}

We plot average dimensionless mass accretion rate $\dot{m}^{\rm
aver}_{\rm inact}$ as a function of power $\kappa_{\rm m}$ of the
accretion rate distribution (see Eq. \ref{mdot_dis}) for the
inactive galaxies in Fig. \ref{fig_kdotminact}. As discussed in \S
2, we can calculate both active and inactive black hole mass
densities as functions of redshift simultaneously using the
bolometric QLF based on the assumption that the growth of massive
black holes is dominated by mass accretion. We plot the total black
hole mass density $\rho_{\rm bh}(z)$, the black hole mass densities
$\rho_{\rm bh}^{\rm inact}(z)$ for inactive galaxies and $\rho_{\rm
bh}^{\rm act}(z)$ for active galaxies, as functions of redshift $z$
in Fig. \ref{fig_zrho}, for different values of $\kappa_{\rm m}$
(i.e., $\dot{m}^{\rm aver}_{\rm act}$), respectively. The ratios of
black hole mass densities $\rho_{\rm bh}^{\rm B}(z)$ accumulated
during ADAF phases between $z$ and $z_{\rm max}$ to the local black
hole mass density $\rho_{\rm bh}^{\rm local}=4\times10^{5}M_{\odot}
{\rm Mpc}^{-3}$ are also plotted in Fig. \ref{fig_zrho}.

\vskip 1.0cm \figurenum{1}
\centerline{\includegraphics[angle=0,width=9.0cm]{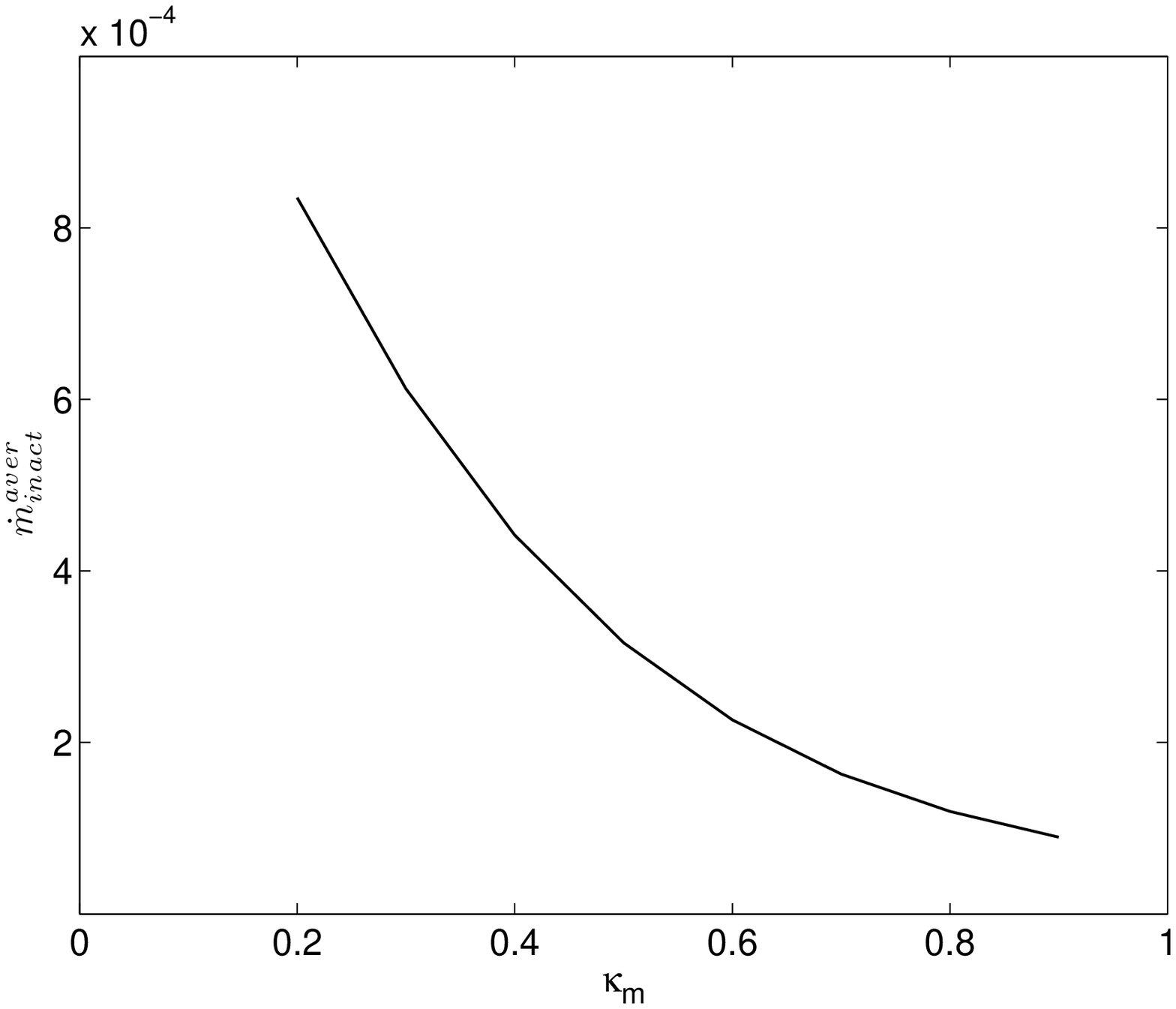}}
\figcaption{The relation of the power $\kappa_{\rm m}$ in Eq.
(\ref{mdot_dis}) with average dimensionless mass accretion rate
$\dot{m}^{\rm aver}_{\rm inact}$ for inactive galaxies. {The minimum
accretion rate, $\dot{m}_{\rm min}=1.0\times10^{-5}$ is adopted in
the calculations.}The relation of the power $\kappa_{\rm m}$ in Eq.
(\ref{mdot_dis}) with average dimensionless mass accretion rate
$\dot{m}^{\rm aver}_{\rm inact}$ for inactive galaxies.  {The
minimum accretion rate, $\dot{m}_{\rm min}=1.0\times10^{-5}$ is
adopted in the calculations.}}\label{fig_kdotminact}

\vskip 1.0cm \figurenum{2}
\centerline{\includegraphics[angle=0,width=9.0cm]{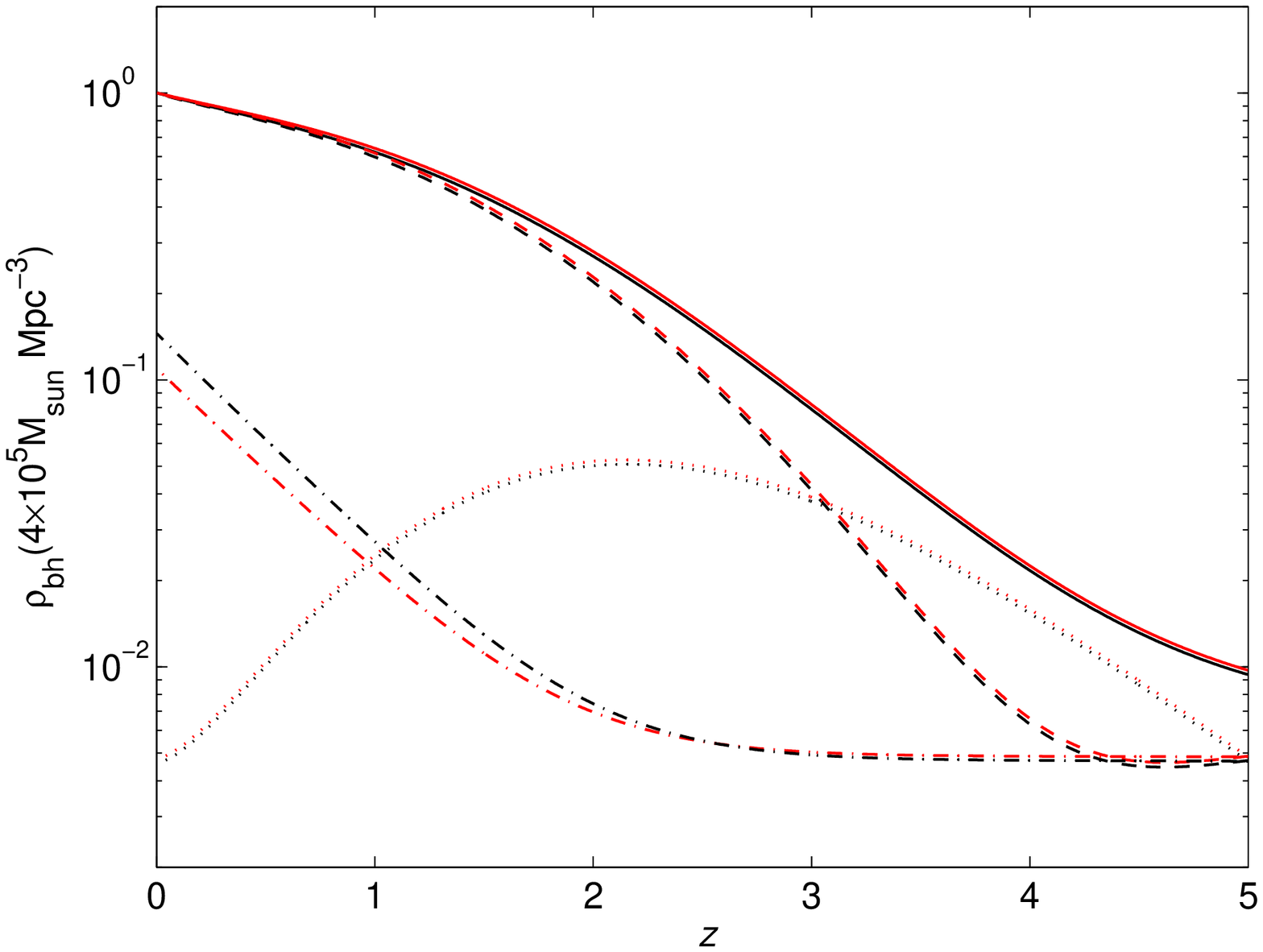}}
\figcaption{The total black hole mass densities $\rho_{\rm bh}(z)$
accumulated through accretion as functions of redshift $z$ (solid
lines) in unit of $\rho_{\rm bh}^{\rm local}=4\times10^{5}M_{\odot}
{\rm Mpc}^{-3}$. The dashed and dotted lines represent the black
hole mass densities $\rho_{\rm bh}^{\rm inact}(z)$ for inactive
galaxies and $\rho_{\rm bh}^{\rm act}(z)$ for active galaxies at
redshift $z$, respectively. The dot-dashed lines represent the ratio
of black hole mass densities $\rho_{\rm bh}^{\rm B}(z)$ accumulated
during ADAF phases to the local black hole mass density $\rho_{\rm
bh}^{\rm local}=4\times10^{5}M_{\odot} {\rm Mpc}^{-3}$ from $z_{\rm
max}$ to $z$. The different colors correspond to different values of
$\kappa_{\rm m}=0.2$ (black) and 0.3 (red).}\label{fig_zrho}

 The global structure of an ADAF surrounding a
spinning massive black hole is available by solving a set of general
relativistic hydrodynamical equations \citep*[see][for the
details]{manmoto00}. The structures of ADAFs are plotted in Fig.
\ref{structure_mdot0p001} for different parameters, which show that
the proton temperature in the inner edge of the ADAFs surrounding
rapidly spinning black holes can be as high as $\sim 10^{12}$K. The
spectra of ADAFs surrounding massive black holes can be calculated
with the derived global structure of ADAFs (see \S 4). In Fig.
\ref{fig_nulnu_delta0p1}, we plot the X-ray and $\gamma$-ray spectra
of the ADAFs surrounding massive black holes with different values
of { spin parameter} $a$. We also plot the spectra calculated with
$\delta=0.01$ in Fig. \ref{fig_nulnu_delta0p01}. In the
calculations, the black hole mass $M_{\rm bh}=10^{8}\rm M_{\odot}$
is adopted.
As the radiative efficiency of an ADAF is no longer constant, we
calculate the bolometric luminosities of ADAFs as functions of mass
accretion rate $\dot{m}$. Figure \ref{fig_lambda_mdot} shows how the
Eddington ratios $L_{\rm bol}/L_{\rm Edd}$ vary with mass accretion
rate $\dot{m}$ for different black hole { spin parameters} and ADAF
parameters. The average X-ray and $\gamma$-ray spectra from a
population of ADAFs, of which the mass accretion rate distribution
is described by Eq. (\ref{mdot_dis}), are given in Fig.
\ref{fig_mean_nulnu} for $\delta=0.1$ and $0.01$, respectively.

The average mass accretion rate $\dot{m}_{\rm inact}^{\rm aver}$ can
be calculated with Eq. (\ref{mdot_aver}) for a specified index
$\kappa_{\rm m}$ of the power law accretion rate distribution
(\ref{mdot_dis}). For a given average mass accretion rate
$\dot{m}_{\rm inact}^{\rm aver}$, the contribution of the ADAFs in
all inactive galaxies to the X-ray/$\gamma$-ray background is
calculated with the derived inactive black hole mass density and
ADAF spectra. We plot the contribution of the ADAFs in all inactive
galaxies to the X-ray band cosmological background with different
black hole and ADAF parameters in Figs. \ref{fig_ele_x_delta0p1} and
\ref{fig_ele_x_delta0p01}. For comparison, we also plot the observed
XRB in the figure, and the sum of the contributions from the type
I/II bright AGNs(Compton-thin) given by \citet{tuv09} and all
inactive galaxies derived in this work. Their contribution to the
extragalactic $\gamma$-ray background is plotted in Figs.
\ref{fig_ele_g_delta0p1} and \ref{fig_ele_g_delta0p01} for
$\delta=0.1$ and $0.01$, respectively.

\vskip 1.0cm \figurenum{3}
\centerline{\includegraphics[angle=0,width=9.0cm]{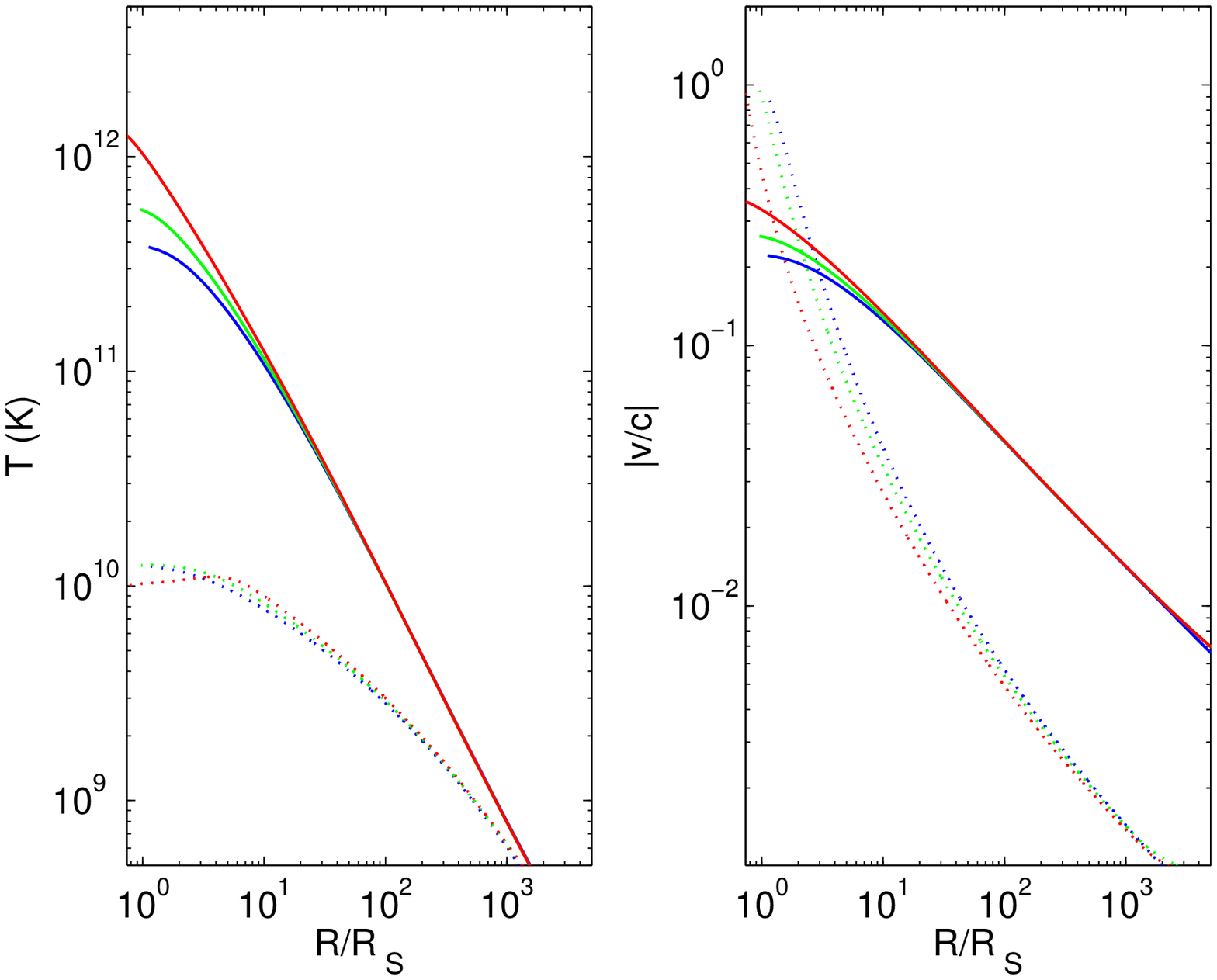}}
\figcaption{The structure profiles of the ADAFs surrounding a
$10^{8}M_{\odot}$ black hole with different spins, $a=0.$ (blue
lines), $a=0.5$(green lines) and $a=0.9$(red lines) accreting at
$\dot{m}=0.001$. Left panel: the temperature distributions of the
electrons (dotted lines) and ions (solid lines) as functions of
dimensionless radius $R/R_{\rm S}$ ($R_{\rm S}=2GM/c^2$). Right
panel: the radial velocity (dotted lines) and sound speed (solid
lines) distributions. The parameter $\delta=0.01$ is adopted in the
calculations. } \label{structure_mdot0p001}

\vskip 1.0cm \figurenum{4}
\centerline{\includegraphics[angle=0,width=9.0cm]{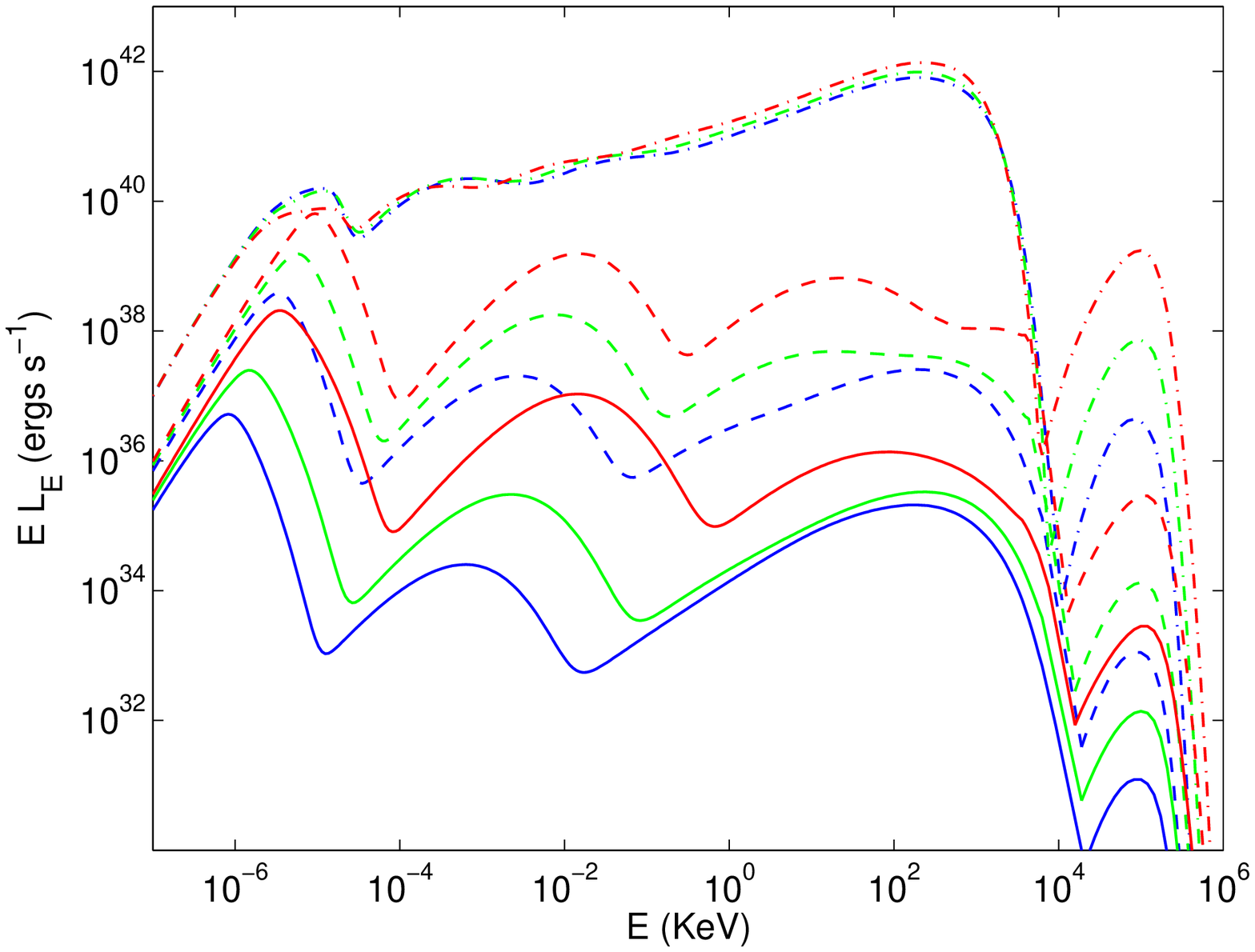}}
\figcaption{The X-ray/$\gamma$-ray spectra of the ADAFs surrounding
a $10^{8}M_{\odot}$  Schwarzschild (blue lines) or Kerr black hole
with $a=0.5$ (green lines) and $a=0.9$ (red lines) accreting at
different rates. The solid, dashed, and dash-dotted lines are
corresponding to the accretion rates $\dot{m}=1.0\times 10^{-5}$,
$1.0\times 10^{-4}$, and $1.0\times 10^{-2}$, respectively. The
parameter $\delta=0.1$ is adopted in the calculations. }
\label{fig_nulnu_delta0p1}

\vskip 1.0cm \figurenum{5}
\centerline{\includegraphics[angle=0,width=9.0cm]{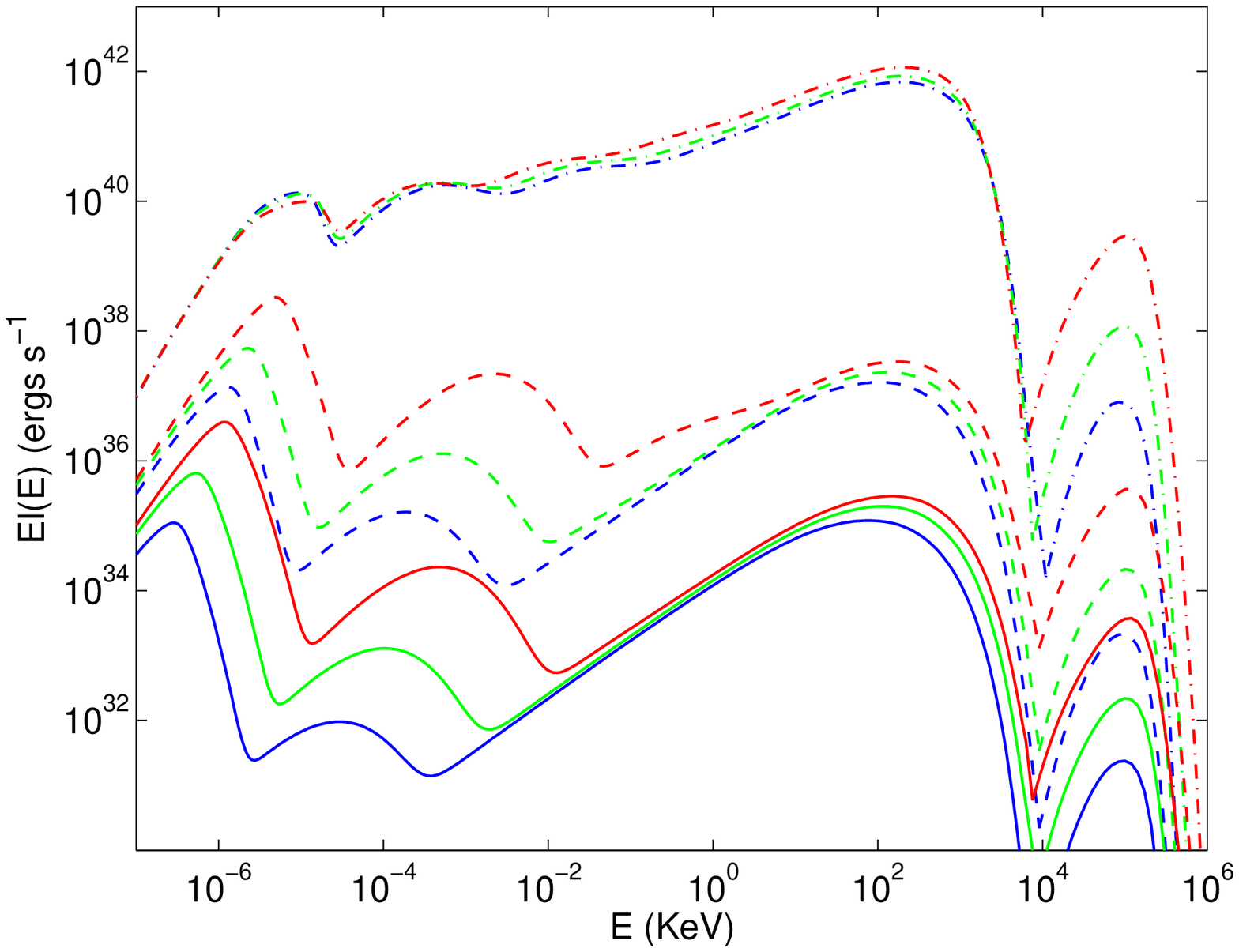}}
\figcaption{The same as Fig. \ref{fig_nulnu_delta0p1}, but the
parameter $\delta=0.01$ is adopted.} \label{fig_nulnu_delta0p01}

\vskip 1.0cm \figurenum{6}
\centerline{\includegraphics[angle=0,width=9.0cm]{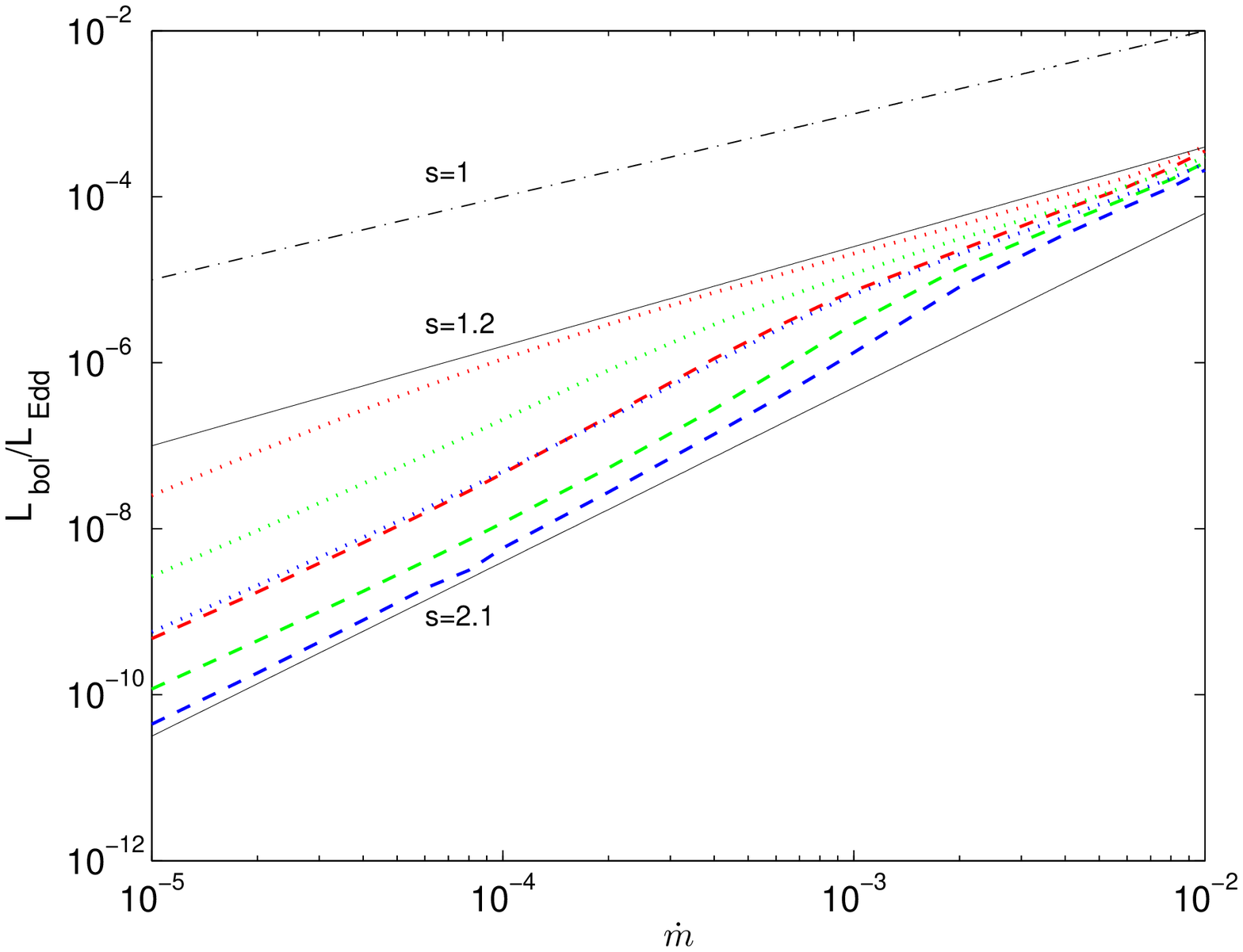}}
\figcaption{The relation between dimensionless mass accretion rate
$\dot{m}$ and Eddington ratio $L_{\rm bol}/L_{\rm Edd}$ for ADAFs in
inactive galaxies. The dotted and dashed lines represent the results
for $\delta=0.1$ and 0.01, respectively. The lines with different
colors correspond to different values of $a=0$ (blue), $0.5$
(green), and $0.9$ (red), respectively. {For comparison, we also
plot several power-law lines $L_{\rm bol}/L_{\rm Edd}\propto
\dot{m}^s$ in the figure. The black dash-dotted line represents
$s=1$, i.e., constant radiative efficiency, and $\epsilon=0.1$. }}
\label{fig_lambda_mdot}

\vskip 1.0cm \figurenum{7}
\centerline{\includegraphics[angle=0,width=9.0cm]{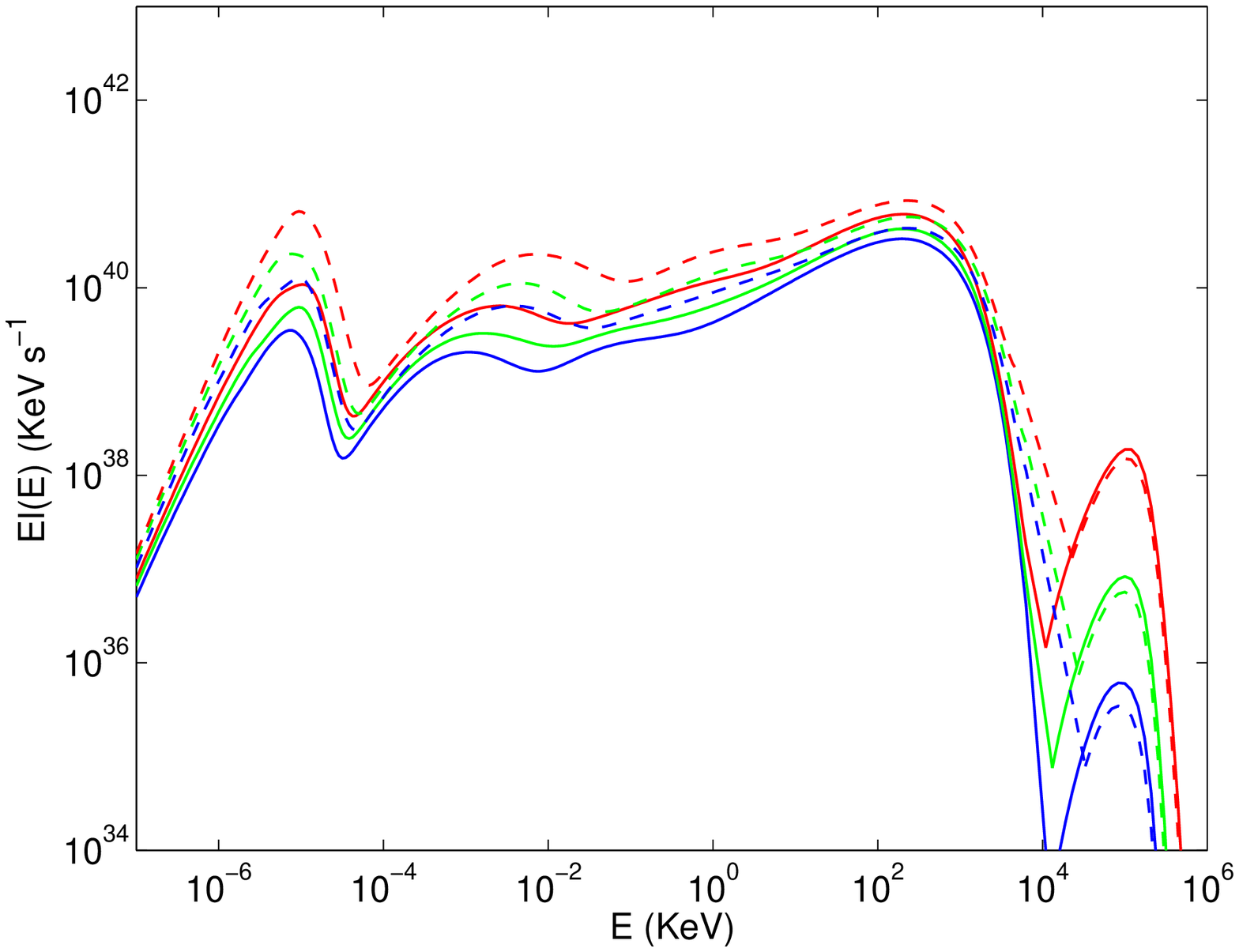}}
\figcaption{The X-ray and $\gamma$-ray spectra averaged over a
population of ADAFs surrounding
 $10^{8}\rm M_{\odot}$ black holes, which has $\dot{m}$ distribution
 as indicated in Eq. \ref{mdot_dis}, with different values of spin
parameter $a$ and ADAF model parameter $\delta$. The different color
lines represent the cases with spin parameter $a=0$ (blue), $0.5$
(green), and $0.9$ (red), respectively. The solid and dashed lines
correspond to the spectra with $\delta=0.1$ and $\delta=0.01$,
respectively. The parameter $\kappa_{\rm m}=0.3$ is adopted in the
calculations.} \label{fig_mean_nulnu}

\vskip 1.0cm \figurenum{8}
\centerline{\includegraphics[angle=0,width=9.0cm]{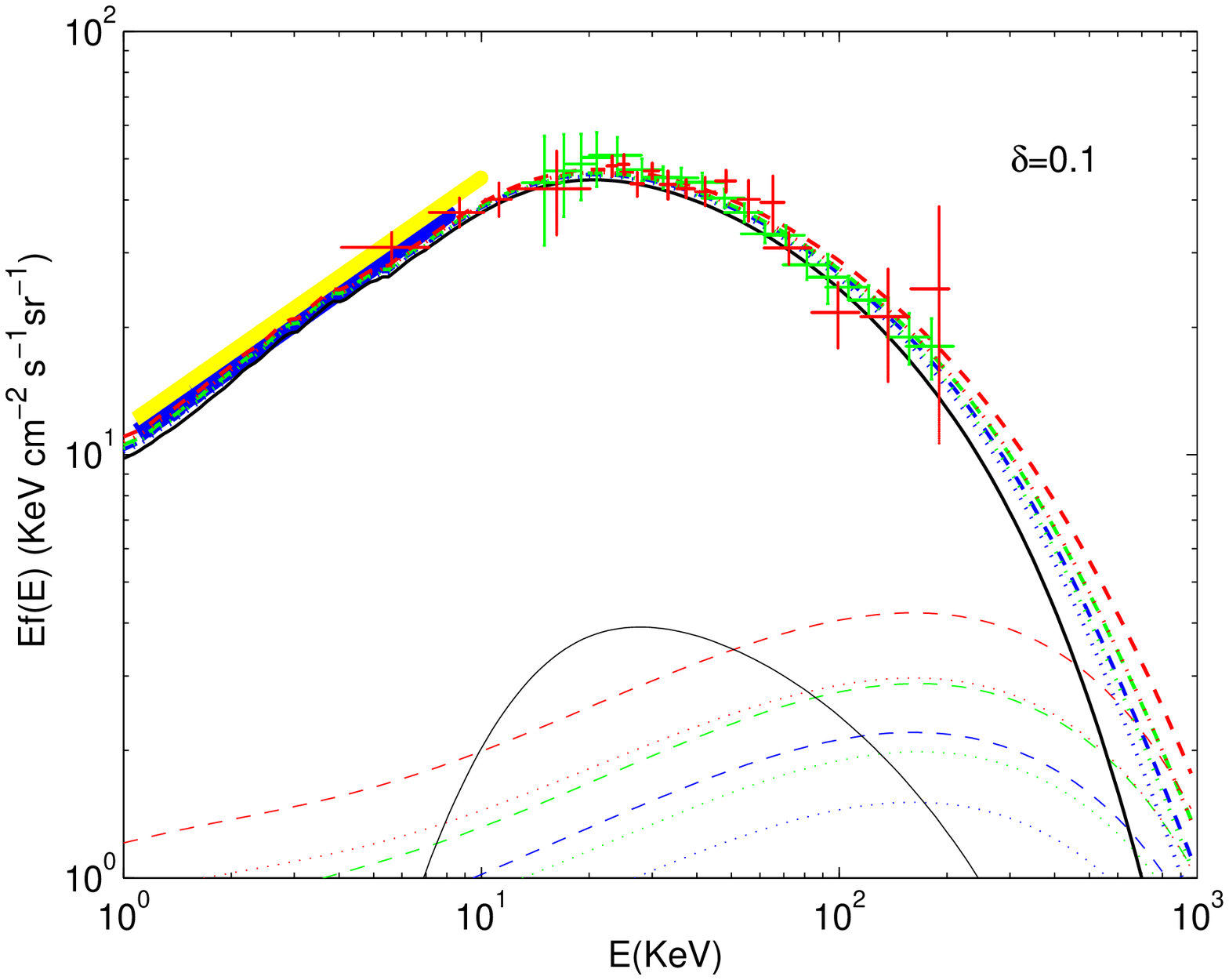}}
\figcaption{The contribution of the ADAFs in all inactive galaxies
to the XRB. The thick black solid line represents the XRB spectrum
from the AGN population synthesis model given by \citet{tuv09}. The
thin solid line represents the contribution to their model from
Compton-thick AGNs. The lower blue, green and red lines correspond
to the contribution from all inactive galaxies with spin parameter
$a=0$, $0.5$, and $0.9$, respectively. The upper blue, green, and
red lines correspond to the sum of X-ray contributions from both the
AGN population synthesis model given by \citet{tuv09} and all
inactive galaxies. The dashed and dotted lines represent different
values of $\kappa_{\rm m}$: 0.2 and 0.3, which correspond to average
mass accretion rate of inactive galaxies $\dot{m}_{\rm inact}^{\rm
aver}=8.35\times10^{-4}$ and $6.12\times10^{-4}$, respectively. The
parameter $\delta=0.1$ is adopted in the calculations. The
measurements of the extragalactic XRB with \textit{Chandra}
\citep{2006ApJ...645...95H}, \textit{XMM}
\citep{2004A&A...419..837D}, \textit{INTEGRAL}
\citep{2007A&A...467..529C} and \textit{Swift} \citep{a08} are
plotted as blue shaded area, yellow shaded area, red  and green
points, respectively. } \label{fig_ele_x_delta0p1}

\vskip 1.0cm \figurenum{9}
\centerline{\includegraphics[angle=0,width=9.0cm]{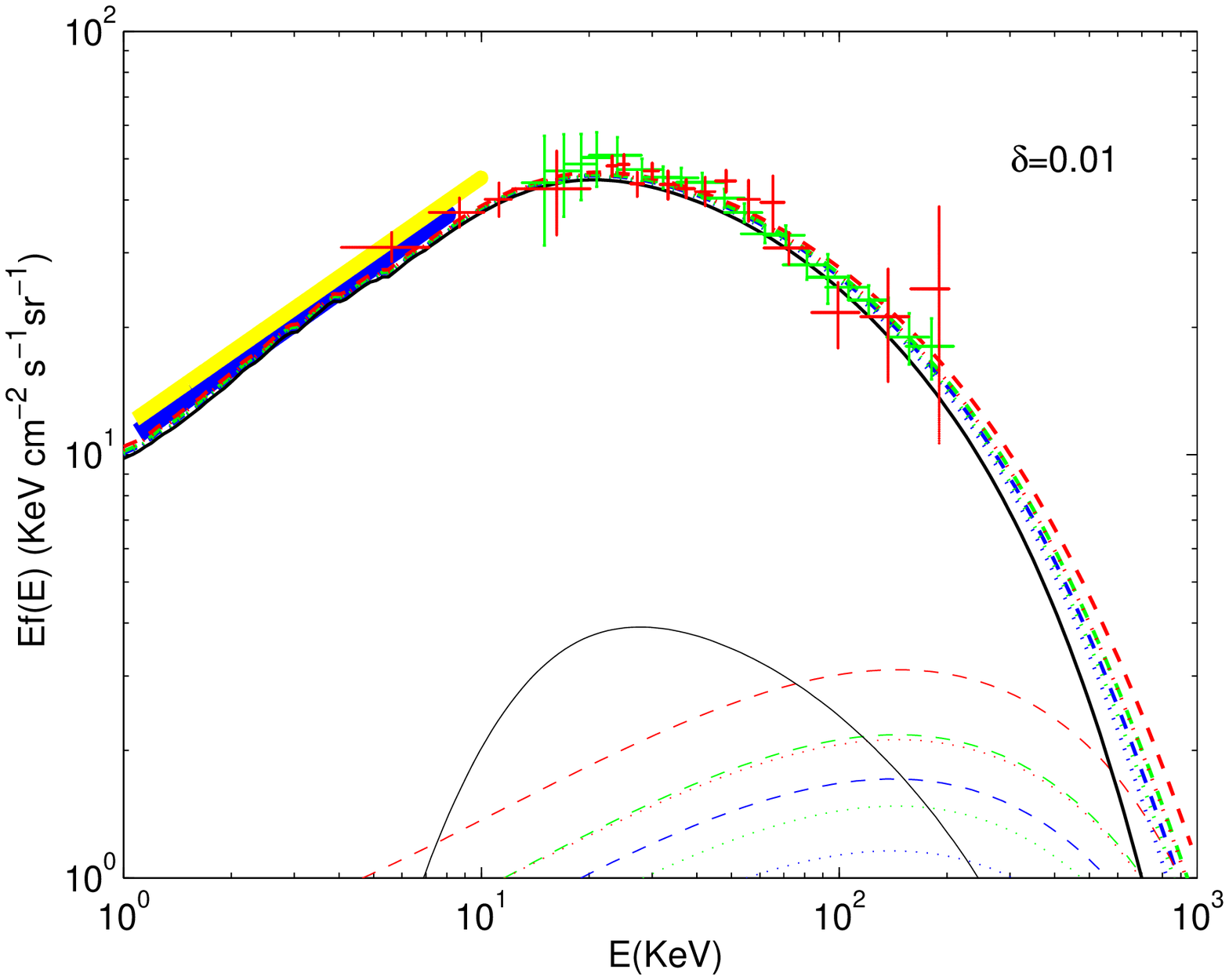}}
\figcaption{The same as Fig. \ref{fig_ele_x_delta0p1}, but the
parameter $\delta=0.01$ is adopted.} \label{fig_ele_x_delta0p01}

\vskip 1.0cm \figurenum{10}
\centerline{\includegraphics[angle=0,width=9.0cm]{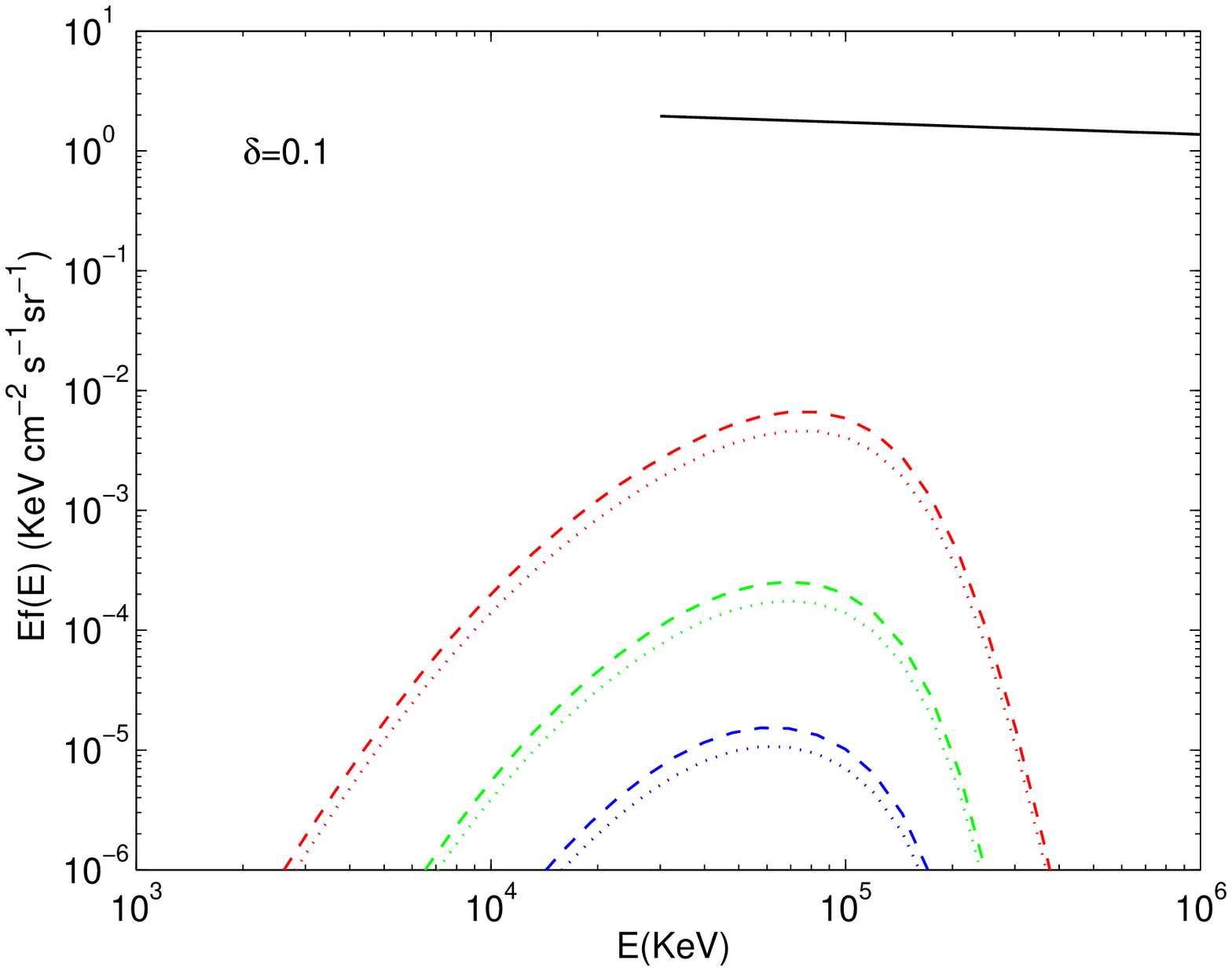}}
\figcaption{The contribution of the ADAFs in all inactive galaxies
to the EGRB. The black solid line denotes the observed EGRB
\citep*[taken from][]{s98}. The different color lines correspond to
the contribution from ADAFs surrounding black holes in all inactive
galaxies with spin parameter $a=0$ (blue), $0.5$ (green), and $0.9$
(red), respectively. } \label{fig_ele_g_delta0p1}

\vskip 1.0cm \figurenum{11}
\centerline{\includegraphics[angle=0,width=9.0cm]{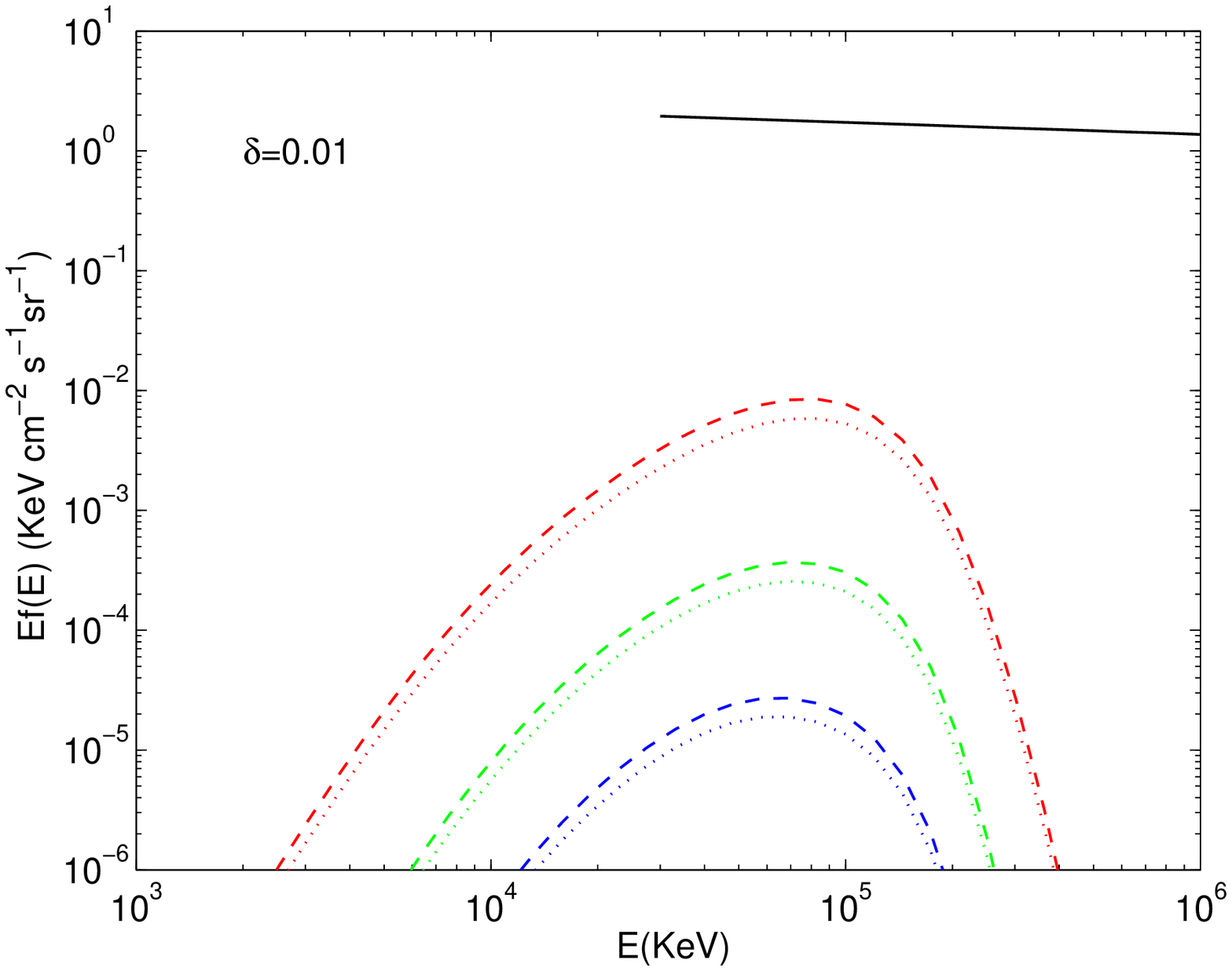}}
\figcaption{The same as Fig. \ref{fig_ele_g_delta0p1}, but the
parameter $\delta=0.01$ is adopted.} \label{fig_ele_g_delta0p01}


\section{Discussion}

Unlike standard thin accretion disks, the radiative efficiency of
ADAFs increases with mass accretion rate $\dot{m}$ \citep{n95}.
Thus, the average radiative efficiency for a population of ADAFs is
available only if the distribution of mass accretion rate $\dot{m}$
is known. The Eddington ratio distribution for AGNs was extensively
explored by different authors \citep*[e.g.,][]{m08,h09,kh09}. These
works suggested that more sources are accreting at lower rates, and
the Eddington ratios for AGNs can be well described by a power law
distribution \citep{h09}.  This implies that the average radiative
efficiency for the ADAFs with this Eddington ratio distribution can
be lower than that for an ADAF accreting at the average mass rate,
because the radiative efficiency of ADAFs increases with mass
accretion rate $\dot{m}$ \citep{n95}. \citet{c07} calculated the
contribution of ADAFs to the XRB, and the constraint on the fraction
of black hole mass accreted in ADAF phases was derived by comparing
with the residual XRB. An average mass accretion rate is adopted in
\citet{c07}'s calculations for simplicity, and therefore the upper
limit on the mass growth of black holes in ADAF phases is
under-estimated. In this work, we improve the calculations by
\citet{c07} by including the Eddington ratio distribution for AGNs
in our calculations.

Using the method described in \S 2, the active, inactive and total
black hole mass densities as functions of redshift are calculated by
tuning the average radiative efficiency $\epsilon$ for active
galaxies to let the total black mass density match the measured
local black hole mass density at $z=0$ (see Fig. \ref{fig_zrho}). It
is found that most of the local black hole mass was accumulated
during $z\la 2$, and fraction of the local black hole mass accreted
during ADAF phases is determined by the average mass accretion rate
$\dot{m}_{\rm inact}^{\rm aver}$, i.e., the distribution of
accretion rates for inactive galaxies. We find that the average
radiative efficiency $\simeq 0.134$ is required for active galaxies,
which corresponds to $a\approx 0.9$. This is consistent with that
derived in \citet{c07} and \citet{e02}.


{In the calculation of the black hole mass densities, we need to
know the value of the average Eddington ratio for the active
galaxies $\lambda_{\rm act}^{\rm aver}$, which is still uncertain.
\citet{md04} estimated that the average accretion rate $\dot{m}^{\rm
aver}_{\rm act}$ varies from 0.1 at $z\sim0.2$ to 0.4 at $z\sim2$
from a large sample of SDSS quasars. \citet{k06} estimated
\texttt{}the black hole masses and Eddington ratios for a sample of
luminous AGNs with $0.3<z<4$, and found that their average Eddington
ratio is $\simeq 0.25$. We adopt $\lambda_{\rm act}^{\rm aver}=0.25$
in this work in the calculation of active black hole mass density
$\rho_{\rm bh}^{\rm act}(z)$ and then the inactive black hole mass
density $\rho_{\rm bh}^{\rm inact}(z)$. The derived active black
hole mass density $\rho_{\rm bh}^{\rm act}(z)$ is proportional to
$1/\lambda_{\rm act}^{\rm aver}$ (see Eq. \ref{bhmdensact}). We find
that the derived $\rho_{\rm bh}^{\rm act}(z)\ll \rho_{\rm bh}^{\rm
inact}(z)$ for $z\la 3$ (see Fig. \ref{fig_zrho}). The inactive
black hole mass density $\rho_{\rm bh}^{\rm inact}(z)$ is calculated
by subtracting the active black hole mass density $\rho_{\rm
bh}^{\rm act}(z)$ from the total black hole mass density $\rho_{\rm
bh}(z)$ at redshift $z$ (see Eq. \ref{rhobhinact}), and $\rho_{\rm
bh}(z)$ is derived with QLF as required to match the measured local
black hole mass density $\rho_{\rm bh}^{\rm local}$ at $z=0$, which
implies that the derived $\rho_{\rm bh}^{\rm inact}(z)$ is almost
insensitive to the value of $\lambda_{\rm act}^{\rm aver}$ for
$z\la3$. The XRB and growth of massive black holes are mainly
contributed by the accretion in the sources at low redshifts, which
implies that our main conclusions on the growth of massive black
holes at their late stage will not be affected if the value of
$\lambda_{\rm act}^{\rm aver}$ does not deviate much from the value
adopted in this work. For the average radiative efficiency
$\epsilon=0.134$ derived in this work, the average mass accretion
rate for active galaxies $\dot{m}_{\rm act}^{\rm aver}=0.187$
corresponding to $\lambda_{\rm act}^{\rm aver}=0.25$. The black hole
mass densities can be calculated alternatively by using the
Eddington ratio distribution given by Eq. \ref{mdot_dis}. This
distribution shows that most active galaxies have mass accretion
rates $\dot{m}$ close to the critical one $\dot{m}_{\rm crit}=0.01$
with an average $\dot{m}_{\rm act}^{\rm aver}\sim 0.1$ for
$\kappa_{\rm m}=0.3$. As discussed above, our main conclusions will
not be altered either with this value or a distribution instead of a
single average one. The contribution of all bright AGNs to the XRB
can be directly calculated with the QLF, which is independent of the
Eddington ratio distribution. \citet{h07}'s calculation showed that
the observed XRB can be roughly reproduced with their QLF. }


Our calculations of the XRB contributed by ADAFs in inactive
galaxies are sensitive to the spectra of ADAFs, which are described
by several parameters, i.e., magnetic field strength relative to gas
pressure $\beta$, the fraction of energy directly heating the
electrons $\delta$, and the viscosity parameter $\alpha$. In
\citet{c07}'s work, the theoretical/observational constraints on the
values of these parameters were discussed, which leads to narrow
ranges for the values of these parameters. The X-ray/$\gamma$-ray
spectra of ADAFs are almost independent of the value of $\beta$
(magnetic field strength), which only affects the spectra in radio
bands. As we are focusing on the hard X-ray and $\gamma$-ray energy
bands, we find that the spectra of ADAFs depend most sensitively on
the value of $\delta$, because the electron temperature in the ADAF
is sensitively affected by the energy directly heating the
electrons. Besides a conventional value of $\delta=0.1$ adopted in
our calculations, we also carry out the calculations with
$\delta=0.01$ for comparison. In this case, the radiative efficiency
of ADAFs is lower than that for $\delta=0.1$, as the heating of
electrons is significantly suppressed (see Figs.
\ref{fig_nulnu_delta0p1}-\ref{fig_lambda_mdot}).

Figure \ref{fig_lambda_mdot} shows how the Eddington ratios $L_{\rm
bol}/L_{\rm Edd}$ vary with mass accretion rate $\dot{m}$ for
different values of $\delta$ and black hole { spin parameter} $a$.
We find that the Eddington ratio varies with $\dot{m}$ roughly as
$L_{\rm bol}/L_{\rm Edd}\propto \dot{m}^s$, where $s\simeq 1.2-2.1$
depending on the values of parameters adopted. {For comparison, we
plot several power-law lines $L_{\rm bol}/L_{\rm Edd}\propto
\dot{m}^s$, $s=1$, i.e., constant radiative efficiency
$\epsilon=0.1$ (black dash-dotted line), $s=1.2$ and $2.1$ in the
same figure.} The result $s\simeq 1.2-2.1$ means that the radiative
efficiencies of ADAFs increases with $\dot{m}$. Our results are
consistent with those adopted in the previous works
\citep{n95,j05,h07}. It is not surprising that the radiative
efficiency increases with $\delta$, and the bolometric luminosity
$L_{\rm bol}$ is higher for rapidly spinning black holes provided
the same accretion rate $\dot{m}$ is adopted.

Combining the derived inactive black hole mass density $\rho_{\rm
bh}^{\rm inact}(z)$ and the average spectra of a population of ADAFs
with the mass accretion rate distribution (Eq. \ref{mdot_dis}), the
contribution of the ADAFs in all inactive galaxies to the XRB can be
calculated. The results are plotted in Figs.
\ref{fig_ele_x_delta0p1} and \ref{fig_ele_x_delta0p01}, which are
compared with the observed XRB. The contribution of bright AGNs to
the XRB was estimated in many previous works
\citep*[e.g.,][]{u03,tu05,g07,tuv09}. We compare the sum of the
contribution of bright AGNs to XRB estimated by \citet{tuv09} and
that of the ADAFs in all inactive galaxies calculated in this work
with the observed XRB. It is found that $\kappa_{\rm m}\ga 0.3$ for
$\delta=0.1$ (or $\kappa_{\rm m}\ga 0.2$ for $\delta=0.01$) is
required in order to let the calculated XRB not surpass the observed
XRB (see Figs. \ref{fig_ele_x_delta0p1} and
\ref{fig_ele_x_delta0p01}). {The dependence of required value of
$\kappa_{\rm m}$ on parameter $\delta$ can be understood that, a
smaller $\delta$ corresponds to a lower radiative efficiency, while
a smaller $\kappa_{\rm m}$ represents a flatter dimensionless mass
accretion rate distribution which has relatively more high-
$\dot{m}(\approx10^{-3}-10^{-2})$ sources. The average mass
accretion rate for inactive black holes is relatively high for a
distribution with a small $\kappa_{\rm m}$, which also corresponds
to a relatively high average radiative efficiency, because the
radiative efficiency of an ADAF increases with $\dot{m}$ (see Fig.
\ref{fig_lambda_mdot}). This implies that the distribution with a
too flat mass accretion distribution (i.e., small $\kappa_{\rm m}$)
will over-produce sources accreting at rates close to $\dot{m}_{\rm
max}$, which will radiates too much to the XRB. On the other hand,
the value of $\kappa_{m}$ can also be constrained by $\kappa$ and
$s$ according to the relation $L_{\rm bol}/L_{\rm Edd}\propto
\dot{m}^s$ and equations (\ref{edd_ratio_dis}) and (\ref{mdot_dis}).
\citet{h09} suggested $\kappa\approx 0.3-0.8$, and our calculation
predicts $s\simeq 1.2-2.1$ for different parameters, then we can
expect $\kappa_{\rm m}\approx \kappa s\approx 0.36-1.67$, which is
roughly consistent with the constraints from comparison with the XRB
in this work. Our results implies that the XRB contributed by
inactive black holes is dominated by the radiation from
high-$\dot{m}$ sources, rather than low-$\dot{m}$ sources. }

{The energy peak of the contribution of the ADAFs in inactive
galaxies to the XRB is around $\sim 100-200$~keV depending on the
values of $\delta$, which accounts for $\sim15-20$\% of the XRB at
these energy peaks (see Figs. \ref{fig_ele_x_delta0p1} and
\ref{fig_ele_x_delta0p01}). The \textit{Swift} measurements on the
XRB have estimated errors of $\sim 3$\% \citep*{a08,tuv09}. Thus, we
suggest that more accurate measurements of the XRB in the energy
band with $\ga 100$~keV in the future may constrain the growth of
massive black holes at their late stage more precisely. We note that
the peak energy of the contribution of Compton-thick AGNs is $\sim
30$~keV, while it is $\sim100-200$~keV for that of ADAFs in inactive
massive black holes. This implies that the constraints on the growth
of massive black holes at their late stage from the XRB have hardly
been affected by the possible uncertainty of the space number
density of Compton-thick AGNs. }

The values of $\kappa_{\rm m}$=0.2 and 0.3 correspond to average
mass accretion rates of inactive galaxies $\dot{m}_{\rm inact}^{\rm
aver}=8.35\times10^{-4}$ and $6.12\times10^{-4}$, respectively (see
Fig. \ref{fig_kdotminact}). This means that the mass fraction of
local black holes grown in ADAF phases should be less than $10.9\%$
(for $\delta=0.1$) and $14.5\%$ (for $\delta=0.01$) (see Fig.
\ref{fig_zrho}). Our result is about twice of that ($\sim 5\%$) for
$\delta=0.1$ given in \citet{c07}. The discrepancy is mainly
attributed to two factors. The first one is that a single average
mass accretion rate adopted in \citet{c07} for the calculations of
ADAF spectra and their contribution to the XRB, which over-estimated
the average radiative efficiency for ADAFs in inactive galaxies, and
therefore under-estimate the fraction of local black hole mass
accreted during ADAF phases. {The another one is that a different
AGN population synthesis model for the newly measured XRB with
\textit{Swift} is adopted in the calculations in this work. Our
present results are roughly consistent with that derived from
observed Eddington ratio distributions in \citet{h06}.}

ADAFs are very hot, and the temperature of the ions in ADAFs can be
as high as $\sim 10^{12}$K (see Fig. \ref{structure_mdot0p001}),
which implies that $\gamma$-ray emission may be produced through the
pion production processes in the proton-proton (p-p) collisions and
subsequently decay of neutral pions \citep{m97}. We calculate their
contribution to the extragalactic $\gamma$-ray background (EGRB),
and find that less than 1\% of the observed EGRB is contributed by
the ADAFs in these faint sources if the massive black holes are
spinning rapidly, while the contribution to the EGRB from ADAFs can
be neglected if the black holes are non-rotating (see Figs.
\ref{fig_ele_g_delta0p1} and \ref{fig_ele_g_delta0p01}). Our results
are consistent with the previous works showing that about $\sim
25\%$ to $\sim 100\%$ of the EGRB can be attributed to the
unresolved blazars \citep*[e.g.,][]{p93,c95,s96,m00,c08,b09}.



\acknowledgments  We thank the anonymous referee for very helpful
comments/suggestions, and Treister E. for providing the data of the
XRB and their AGN population synthesis model. This work is supported
by the NSFC (grants 10778621, 10703003, 10773020, 10821302 and
10833002), the National Basic Research Program of China (grant
2009CB824800), the Science and Technology Commission of Shanghai
Municipality (10XD1405000), and the CAS (grant KJCX2-YW-T03).

{}



\clearpage

\end{document}